\documentclass[conference]{IEEEtran}
\IEEEoverridecommandlockouts

\usepackage{cite}
\usepackage{amsmath,amssymb,amsfonts}
\usepackage{algorithmic}
\usepackage{graphicx}
\usepackage{textcomp}
\usepackage{color}
\usepackage{xcolor}
\usepackage{todonotes}
\usepackage{blindtext}
\usepackage{makecell}
\usepackage{multicol} 
\usepackage{arydshln}
\usepackage{booktabs}
\usepackage{multirow}
\usepackage{listings}
\usepackage{tabularx}
\usepackage{threeparttable}
\usepackage{comment}
\usepackage{amssymb}
\usepackage{enumitem}
\usepackage{balance}

\usepackage{url}

\usepackage{listings, xcolor}

\definecolor{verylightgray}{rgb}{.97,.97,.97}

\lstdefinelanguage{Solidity}{
	keywords=[1]{anonymous, assembly, assert, balance, break, call, callcode, case, catch, class, constant, continue, constructor, contract, debugger, default, delegatecall, delete, do, else, emit, event, experimental, export, external, false, finally, for, function, gas, if, implements, import, in, indexed, instanceof, interface, internal, is, length, library, log0, log1, log2, log3, log4, memory, modifier, new, payable, pragma, private, protected, public, pure, push, require, return, returns, revert, selfdestruct, send, solidity, storage, struct, suicide, super, switch, then, this, throw, transfer, true, try, typeof, using, value, view, while, with, addmod, ecrecover, keccak256, mulmod, ripemd160, sha256, sha3}, 
	keywordstyle=[1]\color{blue}\bfseries,
	keywords=[2]{address, bool, byte, bytes, bytes1, bytes2, bytes3, bytes4, bytes5, bytes6, bytes7, bytes8, bytes9, bytes10, bytes11, bytes12, bytes13, bytes14, bytes15, bytes16, bytes17, bytes18, bytes19, bytes20, bytes21, bytes22, bytes23, bytes24, bytes25, bytes26, bytes27, bytes28, bytes29, bytes30, bytes31, bytes32, enum, int, int8, int16, int24, int32, int40, int48, int56, int64, int72, int80, int88, int96, int104, int112, int120, int128, int136, int144, int152, int160, int168, int176, int184, int192, int200, int208, int216, int224, int232, int240, int248, int256, mapping, string, uint, uint8, uint16, uint24, uint32, uint40, uint48, uint56, uint64, uint72, uint80, uint88, uint96, uint104, uint112, uint120, uint128, uint136, uint144, uint152, uint160, uint168, uint176, uint184, uint192, uint200, uint208, uint216, uint224, uint232, uint240, uint248, uint256, var, void, ether, finney, szabo, wei, days, hours, minutes, seconds, weeks, years},	
	keywordstyle=[2]\color{teal}\bfseries,
	keywords=[3]{block, blockhash, coinbase, difficulty, gaslimit, number, timestamp, msg, data, gas, sender, sig, value, now, tx, gasprice, origin},	
	keywordstyle=[3]\color{violet}\bfseries,
	identifierstyle=\color{black},
	sensitive=false,
	comment=[l]{//},
	morecomment=[s]{/*}{*/},
	commentstyle=\color{gray}\ttfamily,
	stringstyle=\color{red}\ttfamily,
	morestring=[b]',
	morestring=[b]"
}

\definecolor{lightgray}{rgb}{0.95, 0.95, 0.95}
\definecolor{darkgray}{rgb}{0.4, 0.4, 0.4}
\definecolor{purple}{rgb}{0.65, 0.12, 0.82}
\definecolor{editorGray}{rgb}{0.95, 0.95, 0.95}
\definecolor{editorOcher}{rgb}{1, 0.5, 0} 
\definecolor{editorGreen}{rgb}{0, 0.5, 0} 
\definecolor{orange}{rgb}{1,0.45,0.13} 
\definecolor{olive}{rgb}{0.17,0.59,0.20}
\definecolor{brown}{rgb}{0.69,0.31,0.31}
\definecolor{purple}{rgb}{0.38,0.18,0.81}
\definecolor{lightblue}{rgb}{0.1,0.57,0.7}
\definecolor{lightred}{rgb}{1,0.4,0.5}
\definecolor{codegreen}{rgb}{0,0.6,0}
\definecolor{codegray}{rgb}{0.5,0.5,0.5}
\definecolor{codepurple}{rgb}{0.58,0,0.82}
\definecolor{backcolour}{rgb}{0.95,0.95,0.92}

\newcommand{\todoblue}[1]{\todoc{blue}  {[#1]}}

\newcommand{\todocyan}[1]{\todoc{cyan}  {[#1]}}

\newcommand{\todopurple}[1]{\todoc{purple}  {[#1]}}
\newcommand{\todoc}[2]{{\textcolor{#1} {#2}}}
\newcommand{\liulu}[1]{\todopurple{Liulu: #1}}

\newcommand{\yepang}[1]{\todoblue{Yepang: #1}}

\newcommand{\lili}[1]{\todocyan{Lili: #1}}

\def\BibTeX{{\rm B\kern-.05em{\sc i\kern-.025em b}\kern-.08em
    T\kern-.1667em\lower.7ex\hbox{E}\kern-.125emX}}

\begin{document}

\title{Characterizing Transaction-Reverting Statements in Ethereum Smart Contracts}



\author{
    \IEEEauthorblockN{Lu Liu$^{a,b}$, Lili Wei$^{b}$, Wuqi Zhang$^{b}$, Ming Wen$^{c}$, 
    Yepang Liu$^{a*}$, Shing-Chi Cheung$^{b*}$}\thanks{* Yepang Liu and Shing-Chi Cheung are the corresponding authors.}
    \IEEEauthorblockA{$^a$ Department of Computer Science and Engineering, Southern University of Science and Technology, Shenzhen, China}
    \IEEEauthorblockA{$^b$ Department of Computer Science and Engineering, Hong Kong University of Science and Technology, Hong Kong, China}
    \IEEEauthorblockA{$^c$ School of Cyber Science and Engineering, Huazhong University of Science and Technology, Wuhan, China}
    \IEEEauthorblockA{\{lliubf, liliwei, wzhangcb, scc\}@cse.ust.hk, \{mwenaa\}@hust.edu.cn, \{liuyp1\}@sustech.edu.cn
    } 
}


\maketitle

\begin{abstract}
Smart contracts are programs stored on blockchains to execute transactions. 
When input constraints or security properties are violated at runtime, the transaction being executed by a smart contract needs to be reverted to avoid undesirable consequences.
On Ethereum, the most popular blockchain that supports smart contracts, developers can choose among three transaction-reverting statements (i.e., \texttt{require}, \texttt{if...revert}, and \texttt{if...throw}) to handle anomalous transactions.
While these transaction-reverting statements are vital for preventing smart contracts from exhibiting abnormal behaviors or suffering malicious attacks, there is limited understanding of how they are used in practice. 
In this work, we perform the first empirical study to characterize transaction-reverting statements in Ethereum smart contracts. 
We measured the prevalence of these statements in 3,866 verified smart contracts from popular dapps and built a taxonomy of their purposes via manually analyzing 557 transaction-reverting statements.
We also compared template contracts and their corresponding custom contracts to understand how developers customize the use of transaction-reverting statements.
Finally, we analyzed the security impact of transaction-reverting statements by removing them from smart contracts and comparing the mutated contracts against the original ones. 
Our study led to important findings.
For example, we found that transaction-reverting statements are commonly used to perform seven types of authority verifications or validity checks, and missing such statements may compromise the security of smart contracts.
We also found that current smart contract security analyzers cannot effectively handle transaction-reverting statements when detecting security vulnerabilities.
Our findings can shed light on further research in the broad area of smart contract quality assurance and provide practical guidance to smart contract developers on the appropriate use of transaction-reverting statements.

\end{abstract}


\section{Introduction}\label{sec:introduction}

Smart contracts are programs stored on blockchains to execute transactions.
In recent years, smart contracts have been widely used for various purposes such as
to offer financial services~\cite{smartcontractusecase}.
Ethereum~\cite{ethereum_yellow_paper} is the largest decentralized platform for smart contracts with the second biggest blockchain market capitalization~\cite{coinmarketcap}.
There are over one million transactions executed on Ethereum daily~\cite{dailytransaction}.

\begin{figure}[tbp]
\centering
\includegraphics[scale=0.96, trim=2.5cm 16.5cm 4.3cm 0.7cm]{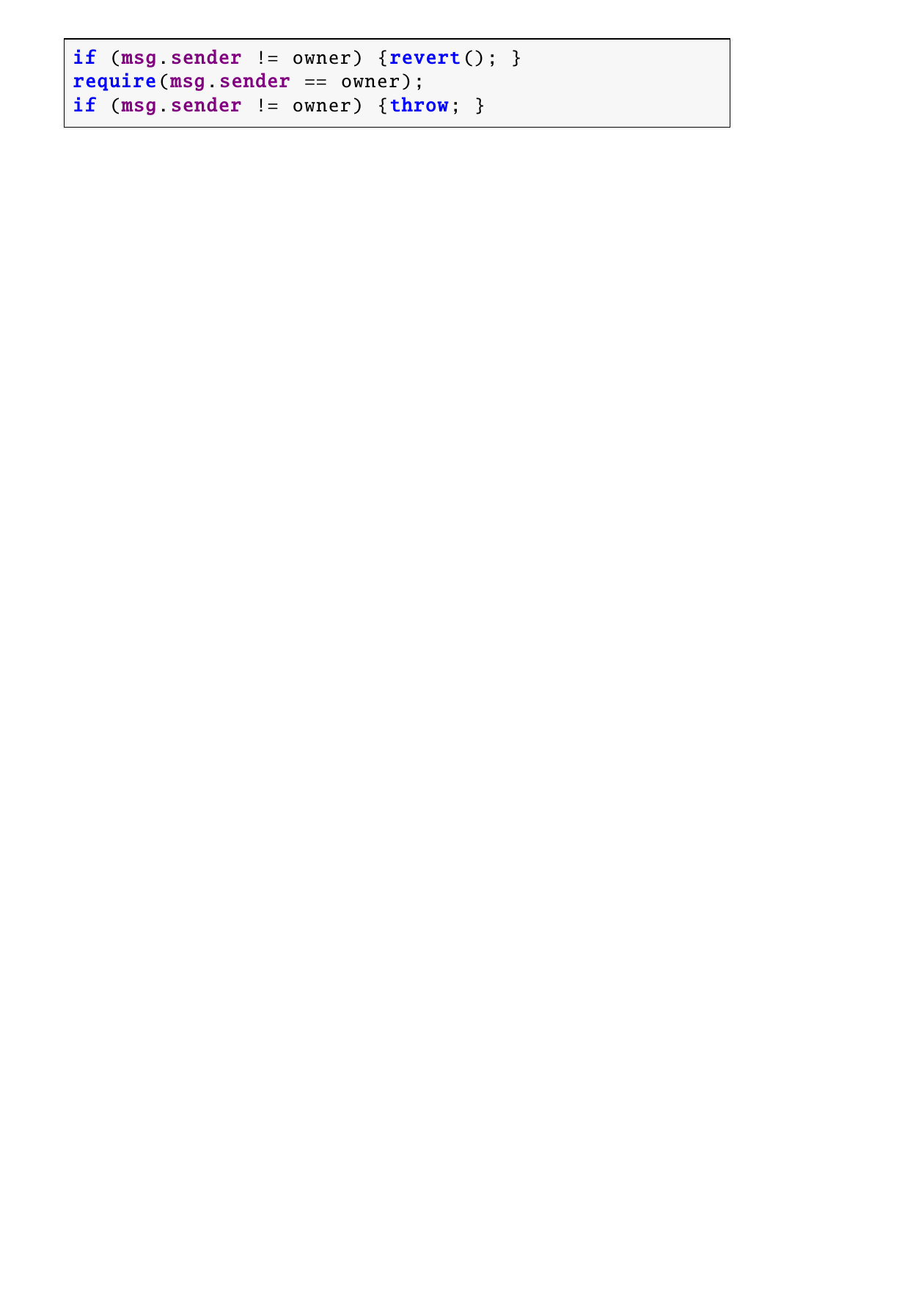}
\caption{Examples of transaction-reverting statements}
\label{transaction-reverting_statements}
\end{figure}

As smart contracts are often used to manage valuable user assets, their security is of paramount importance.
Anomalous transactions caused by various runtime errors should be detected and reverted promptly to prevent undesirable consequences such as financial losses.
In Solidity~\cite{soliditydocumentation}, the most popular programming language for Ethereum smart contracts, there are three statements that can help detect runtime errors and revert transactions, namely, \texttt{require}, \texttt{if...revert}, and \texttt{if...throw}. Figure~\ref{transaction-reverting_statements} shows the example uses of these \textit{transaction-reverting statements} to revert transactions submitted by unauthorized senders.
While all three statements can revert transactions when anomalous conditions occur, the first two would refund the unused gas to transaction senders.

Transaction-reverting statements are frequently used in smart contracts. Our analysis reveals that over 94$\%$ of smart contracts use transaction-reverting statements in certain ways. Surprisingly, this figure is even higher than that of general-purpose \texttt{if} statements. These statements are also frequently discussed in the Solidity developers community. We searched on Stack Overflow\cite{stackoverflow}, the most popular Q\&A website for programmers, using the keywords ``require()'', ``revert()'', and ``if throw'' under the tag ``solidity''. As of August~2021, there are already 1,280 questions related to the three transaction-reverting statements, many of which have been viewed thousands of times. 

\begin{figure}[tb]
    \begin{center}
    \includegraphics[scale=0.95, trim=2.2cm 15.2cm 4.3cm 0cm]{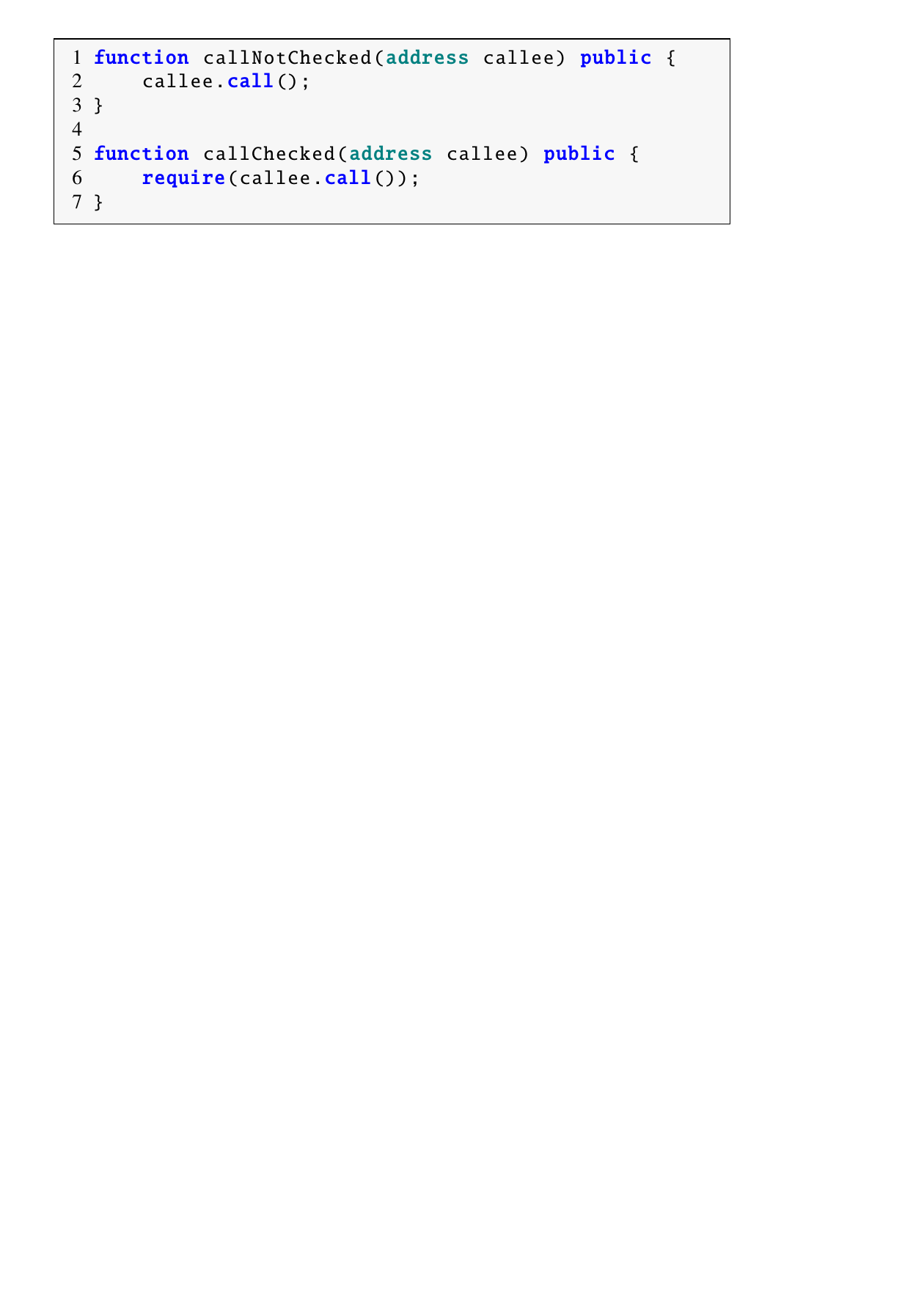}
    \caption{An example of the \textit{Unchecked Call Return Value} weakness}
    \label{swc104}
    \end{center}
\end{figure}

Transaction-reverting statements can effectively help prevent smart contracts from exhibiting abnormal behaviors or suffering malicious attacks.
For example, in the SWC Registry~\cite{swcregistry}, which indexes common smart contract weaknesses, there is a kind of weakness called ``Unchecked Call Return Value'' (SWC-104\cite{swc104}).
This weakness occurs when the return value of a message call is not properly checked in a smart contract. 
To ease understanding, we give an illustrative example in Figure~\ref{swc104}.
In the code snippet, the \texttt{callNotChecked()}function does not check the return value of \texttt{callee.call()} (Line~2).
When the execution of \texttt{callee.call()} fails, the \texttt{callNotChecked()}function would not do anything.
This may cause serious and irreversible consequences, e.g., the contract announces to the caller with error execution information that the call has been executed successfully, but actually, the call fails. 
To fix the weakness, developers are suggested to add a \texttt{require} statement to check the execution status of \texttt{callee.call()} (such as in Line 6 of \texttt{callChecked()})
so that the anomalous transactions can be reverted and the unused gas can be returned to the transaction sender upon unsuccessful execution of \texttt{callee.call()}. 

As we can see from the above example, appropriate uses of transaction-reverting statements can help improve the reliability and security of smart contracts.
However, there is little research on transaction-reverting statements.
Without a comprehensive understanding of how these statements are used in practice, one cannot design tools to effectively identify the inappropriate uses of such statements or formulate good practices to help smart contract developers.
We conducted the first empirical study to characterize transaction-reverting statements in Ethereum smart contracts to bridge the gap. Specifically, we investigated the following four research questions:

\begin{itemize}[leftmargin=1em]
\item \textbf{RQ1 (Prevalence):} \textit{Are transaction-reverting statements commonly used in Ethereum smart contracts?}
\item \textbf{RQ2 (Purpose):} \textit{What are the major purposes of using transaction-reverting statements in smart contracts?}
\item \textbf{RQ3 (Developer Customization):} \textit{Are there differences between template contracts and custom contracts in terms of using transaction-reverting statements?}
\item \textbf{RQ4 (Security Impact):} \textit{Are there any security consequences if transaction-reverting statements are missing in smart contracts?}
\end{itemize}


For the study, we constructed a dataset of 270 template contracts and 3,866
dapp contracts, which were collected from popular template code
repositories~\cite{OpenzeppelinContracts, AragonOS, ConsenSys,
EthereumImprovementProposals} and real-world dapps with millions of
transactions.
To answer RQ1, we measured the code density of transaction-reverting statements and compared it with that of general-purpose \texttt{if} statements in smart contracts.
To answer RQ2, we built a taxonomy of the purposes of transaction-reverting statements via an inductive coding process~\cite{seaman1999qualitative}.
To answer RQ3, we leveraged a code clone detector to identify contracts developers customized from popular-used contract templates and studied how developers customize transaction-reverting statements at a fine granularity of clauses of conditions based on template contracts.
To answer RQ4, we analyzed the security impact of transaction-reverting statements by removing them from smart contracts and comparing the mutated contracts against the original ones.
Our major findings include:

\begin{itemize}[leftmargin=1em]
    \item Over 94$\%$ of our analyzed smart contracts use transaction-reverting statements. Comparatively, only 87.9\% of them use general-purpose \texttt{if} statements. This shows that transaction-reverting statements are pervasively used in real-world smart contracts and may play important roles in assuring the correct execution of transactions.
    \item Transaction-reverting statements are commonly used to perform seven types of security-critical checks, such as verifying user authorities.
    \item  Developers are most likely to strengthen transaction-reverting statements by adding clauses, variables, or new transaction-reverting statements. The customized transaction-reverting statements are commonly used for range checks and logic checks.
    \item The lack of transaction-reverting statements may introduce security issues to smart contracts. Existing smart contract security analyzers show weak support in handling transaction-reverting statements when detecting security vulnerabilities.
\end{itemize}

To summarize, the main contribution of this work is the character study of transaction-reverting statements in Ethereum smart contracts. 
To the best of our knowledge, this study is the first of its kind.
The findings can facilitate further research in smart contract quality
assurance and provide practical guidance to smart contract developers on the
appropriate use of transaction-reverting statements. Our data are released on GitHub for public usage~\cite{dataset}.

The organization of the remaining sections is as follows. 
In Section~\ref{sec:background}, we introduce some related background knowledge. 
Section~\ref{sec:data_collection} presents how we constructed four datasets of smart contracts for empirical analysis. 
Then in Section~\ref{sec:empirical-study}, we present the design of the empirical study to answer the four research questions and introduce our data analysis methodologies and empirical findings. 
We discuss threats to the validity of our studies in Section~\ref{sec:threats}.
After that, we discuss related works in Section~\ref{sec:related_work} and conclude our work in Section~\ref{sec:conclusion}.

\section{Background}\label{sec:background}
This section presents the background and explains the terminologies used in the paper.
\subsection{Smart Contracts \& Dapps}
Smart contracts are autonomous programs running on blockchains like Ethereum~\cite{ethereum_yellow_paper}. 
The execution of smart contracts does not rely on a trusted third party and is fully decentralized.
Dapps are decentralized applications that can offer end-users various functionalities.
The core logic of dapps is backed by smart contracts to meet the requirements of applications.
Solidity~\cite{soliditydocumentation} is the most popular high-level language to program Ethereum smart contracts.
In this paper, we focus on the smart contracts written in Solidity.

\subsection{Error-Handling Statements}\label{ssec:bg-error-handling-statements}
Solidity uses state-reverting exceptions to handle errors. It provides four statements to deal with errors, namely, \texttt{require}, \texttt{if...revert}, \texttt{assert}, and \texttt{if...throw}.
If these statements identify the occurrence of erroneous conditions, they will throw an exception and revert the blockchain and contract state to the state before the execution of the transaction.
The four error-handling statements can be further divided into the following two categories~\cite{soliditydocumentation}:

\begin{itemize}[leftmargin=1em]

\item \textbf{Transaction-reverting statements} refer to the \texttt{require}, \texttt{if...revert}, and \texttt{if...throw} statements, that are used to check for erroneous conditions.
Before version~0.4.10, Solidity provides the \texttt{if...throw} statement for reverting transactions. 
As the language evolves, there are two more alternatives, namely, \texttt{require} and \texttt{if...revert}, to replace \texttt{if...throw} since Solidity~0.4.10.
The \texttt{if...throw} statement is officially deprecated in Solidity~0.4.13. 
These statements can all trigger state reversion when erroneous conditions occur.
The only difference between \texttt{if...throw} and the two replacements is that \texttt{if...throw} will use up all remaining gas when errors occur, while the two replacements will refund the remaining gas to the transaction sender.

\item \textbf{The assertion statement} \texttt{assert} should only be used for debugging purpose, which are not supposed to exist in production code.
If a specified assertion is violated, it means that the contract has a bug, which needs to be fixed. 

\end{itemize}

In our study, we focus on transaction-reverting statements. Since \texttt{if...throw} statement is already deprecated, we mainly investigate the use of \texttt{require} and \texttt{if...revert} statements in real-world smart contracts. In the following of this paper, transaction-reverting statements refer to \texttt{require} and \texttt{if...revert} statements if not otherwise specified.

\subsection{Template Contracts \& Custom Contracts}\label{ssec:bg-template-contracts}


Writing a smart contract is non-trivial for developers, especially when there is a high demand for security~\cite{contract_development}. 
To facilitate contract development and prevent vulnerabilities, \textit{template contracts} are provided by industrial institutions and organizations for different use cases.
These template contracts are usually well-maintained and provide many high-quality or fully functional components for reuse. 
In practice, many developers copy or reuse components in template contracts in their own contracts, which we call \textit{custom contracts}, to save efforts and ensure security. 
Developers' customizations may add, delete, or modify existing transaction-reverting statements for various purposes.

\subsection{Solidity Components} 
Smart contracts written in Solidity are put in \texttt{.sol} files, each of which may contain one or more components of three kinds: \textit{contracts}, \textit{libraries}, and \textit{interfaces}. 
Template contract codebases often provide a set of such Solidity components that developers can reuse. 

\section{Dataset Construction} \label{sec:data_collection}
To investigate our research questions, we constructed four datasets of smart contracts for empirical analysis.
This section explains how these datasets were constructed.
\subsection{Crawling Dapp Contracts}
As of April 2021, over 40 million smart contracts have been deployed on Ethereum~\cite{bigquery_ethereum}. 
Despite the large volume, many Ethereum smart contracts are deprecated or rarely used (with few transactions). 
In our empirical study, we aim to analyze representative smart contracts that are often used in real life.
For this purpose, we chose to collect smart contracts from popular dapps.
Such collected smart contracts are of higher quality, more frequently used, and better maintained.

Specifically, we collected smart contracts from all 1,699 dapps indexed by Dapp.com~\cite{Dappcom}, a popular dapp collection website, in February 2021 by referring to the contract addresses listed in the description of dapps. 
We found that most dapps have less than 200 contracts, but the dapp Uniswap is an exception.
Uniswap~\cite{uniswap} is a decentralized exchange that allows users to exchange one kind of token for another kind. 
It has 3,964 smart contracts because there is a contract factory, which will create a contract for every directly exchangeable token pair on Uniswap, and most such created contracts share the same code.
To reduce the impact of data imbalance, we randomly selected 200 contracts for Uniswap (i.e., downsampling). 
For the other dapps, we collected all  the addresses of their used smart contracts listed on Dapp.com. 
Then, we leveraged the APIs provided by Etherscan~\cite{etherscan}, an Ethereum block explorer, to collect contract source codes.
In total, we collected 6,016 smart contracts, and 3,866 of them are verified ones with source code available, which will be used in the subsequent studies.
Table~\ref{dapp_dataset} provides the demographic information of the 3,866 verified contracts.
As we can see, they are from different categories, contain hundreds of lines of code (on average), and have a large number of transactions.


\begin{table}[]
	
\setlength{\tabcolsep}{4pt}

\caption{Information of The Dapp Contract Dataset
}
\label{dapp_dataset}

\centering
\begin{tabular}{lrrrr}
\hline
\textbf{Category}     & \begin{tabular}[r]{@{}l@{}}\textbf{\# Contracts}\end{tabular} & \textbf{Total LOC} & \textbf{Avg LOC} & \begin{tabular}[r]{@{}l@{}}\textbf{\# Transactions}\end{tabular} \\ 
\hline
Exchanges    & 371             & 256,652   & 695.53  &  1,001,698                  \\ 
Finance      & 621             & 460,218   & 742.29  &  1,257,473                 \\ 
Gambling     & 844             & 864,530   & 1,018.29 &  712,299                  \\ 
Game         & 1,152           & 853,524   & 708.32  &  2,028,069                 \\ 
High-risk    & 446             & 265,960   & 593.66  &  782,583                 \\ 
Marketplaces & 140             & 120,620   & 815.00  &  395,204                 \\ 
Social       & 70              & 45,146    & 654.29  &  152,429                 \\ 
Utilities    & 132             & 78,737    & 601.05  &  376,976                 \\ 
Others       & 90              & 55,950    & 608.15  &  80,302                 \\ 
\hline
Total        & 3,866           & 3,001,337 & 763.50  &  6,787,033                 \\ 
\hline
\end{tabular}%
\end{table}


\subsection{Collecting Template Contracts}
Template contracts play an important role in the Ethereum ecosystem.
When reusing them, developers may customize the transaction-reverting statements. 
To study such customizations, we built a dataset of custom contracts and the corresponding template contracts.
For template contracts, we collected them from four data sources, which contain smart contracts that are widely used on Ethereum. 
For custom contracts, we explain how to identify them in the next subsection.
Table~\ref{template_codebase_info} shows the popularity of the data sources of template contracts, and we introduce each of them in the following. 

\begin{table}[]

\caption{The Popularity of Template Contract Data Sources}
\label{template_codebase_info}

\centering

\begin{tabular}{lrr}
\hline
\textbf{Data Source}   & \textbf{\# GitHub Stars} & \textbf{\# Repository Forks} \\ \hline
OpenZeppelin & 10k     & 4.5k    \\ 
aragonOS     & 488     & 190     \\ 
ConsenSys    & 4.1k    & 774     \\ 
EIPs         & 6.2k    & 2.4k    \\ \hline
\end{tabular}
\end{table}

\begin{itemize}[leftmargin=1em]
\item OpenZeppelin~\cite{OpenzeppelinContracts} is a library for secure smart contract development, which provides reusable contract templates such as implementations of token standards to help build custom contracts. We collected 115 contracts from OpenZeppelin.


\item aragonOS\cite{AragonOS} is a smart contract framework for building decentralized organizations, dapps, and protocols. We collected 107 contracts from aragonOS.

\item ConsenSys\cite{ConsenSys} provides Solidity smart contract code for simple, standards-compliant tokens on Ethereum. We collected 34 contracts from ConsenSys. 

\item Besides the above data sources, we also collected 14 final EIPs (\underline{E}thereum \underline{I}mprovement \underline{P}roposals) with 10 reusable template contracts from the ERC website~\cite{EthereumImprovementProposals}.
\end{itemize}


In total, we collected 270 template contracts. 

\subsection{Identifying Custom Contracts}\label{subsec:custom-contracts}
It isn't easy to associate template contracts with custom contracts because developers rarely explicitly specify the templates they reuse to write smart contracts.
To identify custom contracts, we leveraged a code clone detection tool, SmartEmbed~\cite{gao2020checking}, to calculate the code similarity between the 270 template contracts and our collected 3,866 dapp contracts.
If the code similarity between a dapp contract and a template contract is higher than or equal to 85\%, we consider that the dapp contract is a custom contract based on the template contract.
Via this process, we identified a set of 227 custom contracts based on 74 template contracts.
We give more details of the custom contract dataset in Section~\ref{subsec:rq3}.

\subsection{Creating Mutated Contracts}
To investigate the security impact of transaction-reverting statements in smart contracts, we constructed a dataset of mutated contracts by removing all transaction-reverting statements in the 3,866 contracts.
The mutated contracts were later analyzed by existing smart contract vulnerability detection tools to assess their security.
A detailed description of the mutated contracts is given in Section~\ref{subsec:rq4}.

\section{Empirical Study}\label{sec:empirical-study}

With the four datasets, we conducted a large-scale empirical study, aiming to 1) understand the use of transaction-reverting statements in smart contracts, 2) identify good/bad practices, and provide suggestions to help developers appropriately use transaction-reverting statements, and 3) inspire future research.
In this section, we present our data analysis methodology and empirical findings for each of the four research questions listed in Section~\ref{sec:introduction}.  


\subsection{RQ1 (Prevalence)}

\textbf{Study Methodology:}
To answer RQ1, we measured the prevalence of transaction-reverting statements in smart contracts.
Specifically, we first identified all the transaction-reverting statements in the 3,866 dapp contracts and then computed the code density of these statements.
Following existing practices~\cite{yuan2012characterizing,harty2021logging}, we computed code density for transaction-reverting statements as LOC/LOT, where LOC is the lines of code of a contract and LOT is the lines of transaction-reverting statements.
Similarly, we also computed the code density for general-purpose \texttt{if} statements and \texttt{if...throw} statements for comparison.
Note that we separately analyzed general-purpose \texttt{if}, \texttt{if..throw}, and \texttt{if...revert} statements.
When an \texttt{if} statement is used with \texttt{throw} or \texttt{revert}, we will not consider it as a general-purpose \texttt{if} statement since it is used to revert transactions.
In addition, we counted the number of transaction-reverting statements within a \texttt{if...throw} or \texttt{if...revert} code block as one.




\vspace{0.5em}
\textbf{Finding 1: }\textit{In our analyzed smart contracts, transaction-reverting statements are more frequently used than general-purpose \texttt{if} statements.}
\vspace{0.5em}

Among all the 3,866 contracts, 3,647 (94.3\%) contracts contain transaction-reverting statements, while only 3,399 (87.9\%) contracts contain general-purpose \texttt{if} statements. 
Table~\ref{code_density} gives the detailed results, where the column \textit{``Total Lines of Statements''} lists the total number of the concerned statements in the whole dataset and the \textit{``Code Density''} column shows the average code density per contract for each type of statement.
As shown in the table, transaction-reverting statements are more frequently used than general-purpose \texttt{if} statements in terms of both metrics.
On average, there is one transaction-reverting statement per 49.76 lines of code, while general-purpose \texttt{if} statements are used once per 86.92 lines.

\begin{table}[]

\caption{Code Density of Conditional Statements}
\label{code_density}

\begin{tabular}{lrr}
\hline
\textbf{Conditional Statement Type}      & \begin{tabular}[r]{@{}l@{}}\textbf{Total Lines}\\ \textbf{of Statements}\end{tabular} & \textbf{Code Density} \\ \hline
Transaction-reverting Statement & 67,770                                                              & 49.76        \\ 
\texttt{if...throw} Statement                 & 2,061                                                               & 286.74       \\
General-purpose \texttt{if} Statement                    & 66,122                                                              & 86.92        \\ \hline
\end{tabular}
\end{table}



\vspace{0.5em}
\textbf{Finding 2: }\textit{8.6\% of our analyzed smart contracts are still using the deprecated \texttt{if...throw} statements, which may cause unnecessary financial loss to users.}
\vspace{0.5em}

As explained in Section~\ref{ssec:bg-error-handling-statements}, \texttt{if..throw} statements can also help revert transactions but using them would incur additional costs of gas and induce unnecessary financial loss to the contract users.
As a result, \texttt{require} and \texttt{if...revert} statements were introduced in Solidity~0.4.10 as replacements and \texttt{if..throw} was officially deprecated since Solidity~0.4.13 in 2017.
However, we found that 332 (8.6\%) of our analyzed smart contracts are still using \texttt{if...throw} statements. 
Besides, in 252 smart contracts (6.5\%), there exists a mixed use of \texttt{if...throw} and \texttt{require} statements.
We further collected the Solidity versions used in these 3,866 contracts.
Our results showed that 43 contracts (1.1$\%$) still use Solidity versions before 0.4.10.
Such contracts can only use the deprecated \texttt{if...throw} statements to revert transactions.
The users who submit transactions to these contracts may suffer from unnecessary costs of gas.



\vspace{0.5em}
\noindent 
\setlength{\fboxsep}{0.5em} 
\fbox{\parbox{0.95\linewidth}{ 
\textbf{Answer to RQ1:} 
\textit{Transaction-reverting statements are more frequently used in smart contracts than general-purpose \texttt{if} statements. There are still a non-negligible proportion of contracts using deprecated \texttt{if..throw} statements, which may incur unnecessary gas consumption when transactions revert.}~

\vspace{0.5em}
\textbf{Implication:}
\textit{Transaction-reverting statements may play an essential role in assuring the correct execution of transactions.
Researchers working on smart contract quality assurance and security analysis should pay more attention to such statements as inappropriately using them may lead to abnormal contract behaviors or financial losses.
}
}}
\vspace{0.5em}

\subsection{RQ2 (Purpose)}\label{subsec:purpose taxonomy}

\textbf{Study Methodology:}
To understand the purposes of using transaction-reverting statements, we manually analyzed our collected smart contracts with the following two steps:

\textbf{Step 1: Statement selection.} Since there are 67,770 transaction-reverting statements in the 3,866 dapp contracts, it is infeasible to analyze all of them manually.
For our study, we randomly selected 382 of these statements, representing the whole set with a confidence level of 95$\%$ and a confidence interval of 5$\%$.
For the 270 template contracts, we analyzed all 175 transaction-reverting statements in them.

\textbf{Step 2: Constructing the purpose taxonomy.} To understand and categorize the purposes, we first sampled 100 of the 557~(=~382~+~175) transaction-reverting statements for a pilot construction of the taxonomy. Similar to many existing empirical studies, we followed an open coding procedure~\cite{seaman1999qualitative} to inductively create the categories of our taxonomy in a bottom-up manner. 
Two authors read all the sampled transaction-reverting statements and the corresponding contracts to understand their purposes. The two authors also considered the string arguments of the transaction-reverting statements provided by the contract owners and the comments around the transaction-reverting statements when comprehending the contract code. They categorized the 100 statements independently and marked those unclear or insufficient categories. They then discussed and adjusted their category tags during meetings with the help of a third author to resolve conflicts. In this way, we successfully constructed the pilot taxonomy. 

Based on the coding schema in the pilot taxonomy, the two authors continued to label the remaining 457 transaction-reverting statements for two more iterations. In these two iterations, the two authors went back and forth between categories and transaction-reverting statements to refine the taxonomy. The conflicts of labeling were again discussed during meetings and resolved by the third author. In this way, we adjusted the pilot taxonomy and obtained the final results. 
We used the Cohen’s Kappa score~\cite{cohen1960coefficient} to measure the agreement between the two authors. The overall score is 0.73, indicating that the two authors had a high agreement on the taxonomy.


As shown in Table~\ref{purpose of use}, the final taxonomy is organized into two categories, each of which is further divided into sub-categories. There is no overlap between these sub-categories, i.e., a clause in a transaction-reverting statement can only be classified into one of them. 
The table also provides illustrative examples collected from our datasets to ease understanding. 
\begin{table*}[]
    \centering
    \caption{The Purposes of Using Transaction-Reverting Statements in the paragraphs.}
    \label{purpose of use}
    \resizebox{\textwidth}{!}{%
    \begin{tabular}{|l|l|l|l|p{6cm}|}
    \hline
    \textbf{Category}                                                                           & \begin{tabular}[c]{@{}l@{}}\textbf{First-Level}\\ \textbf{Sub-Category}\end{tabular}                   & \begin{tabular}[c]{@{}l@{}}\textbf{Second-Level}\\ \textbf{Sub-Category}\end{tabular}          & \textbf{Description}                                                                                                                                                                      & \textbf{Illustrative Example} \\ \hline
    \multirow{3}{*}{\begin{tabular}[c]{@{}l@{}}\\ \\ \\ Authority \\ Verification\end{tabular}} & \multirow{2}{*}{\begin{tabular}[c]{@{}l@{}}Address\\ Authority\\ Check\end{tabular}} & \begin{tabular}[c]{@{}l@{}}Equal to \\ a specific \\ address\end{tabular}    & \begin{tabular}[c]{@{}l@{}}Check whether a contract address \\ is equal to a 20 bytes address \\ specified by the contract owner.\end{tabular}                                                & \begin{tabular}[c]{@{}l@{}} \lstinline|require(msg.sender == | \\ \lstinline|address(nonFungibleContract));| \end{tabular} \\ \cline{3-5} 
                                                                                       &                                                                                      & \begin{tabular}[c]{@{}l@{}}Within \\ a specific \\ address list\end{tabular} & \begin{tabular}[c]{@{}l@{}}Check whether a contract address \\ is in a address list provided by \\ the contract owner. \end{tabular} & \begin{tabular}[c]{@{}l@{}} \lstinline|require(isAuthorized(msg.sender, | \\ \lstinline|msg.sig)); | \end{tabular} \\ \cline{2-5} 
                                                                                       & \begin{tabular}[c]{@{}l@{}}Token \\ Verification\end{tabular}                        & -                                                                            & \begin{tabular}[c]{@{}l@{}}Some developers use a token ID \\ to identify a contract address. \\ The verification of the token ID \\ is actually the verification of \\ a contract address.\end{tabular}        & \begin{tabular}[c]{@{}l@{}} \lstinline|require(_exists(tokenId), |\\ \lstinline|"URI query for nonexistent token");| \\ (\textit{$\_$exists()} maps a token ID to a contract address \\ using a mapping to verify the contract address.) \end{tabular} \\ \hline
    \multirow{5}{*}{\begin{tabular}[c]{@{}l@{}}\\ \\ \\ \\ \\ Validity\\ Check\end{tabular}}          & \begin{tabular}[c]{@{}l@{}}Logic\\ Check\end{tabular}                                & -                                                                            & \begin{tabular}[c]{@{}l@{}}Check runtime values\\ using logical operators.\end{tabular}                                                                                               & \begin{tabular}[c]{@{}l@{}} \lstinline|require(!_mintingFinished);| \end{tabular} \\ \cline{2-5} 
                                                                                       & \begin{tabular}[c]{@{}l@{}}Range\\ Check\end{tabular}                                & -                                                                            & \begin{tabular}[c]{@{}l@{}}Check whether the value of \\ a variable is within a value range.\end{tabular}                                                                                 & \begin{tabular}[c]{@{}l@{}} \lstinline|require(underlyingBalance > 0, | \\ \lstinline|"Not have any liquidity deposit");| \end{tabular} \\ \cline{2-5} 
                                                                                       & \begin{tabular}[c]{@{}l@{}}Overflow/\\ Underflow\\ Check\end{tabular}                & -                                                                            & \begin{tabular}[c]{@{}l@{}}
                                                                                        Check whether the value of the \\ variable is out of range of the \\ declared data type.
                                                                                    \end{tabular} & \begin{tabular}[c]{@{}l@{}} \lstinline|require((z = x + y) >= x);|\end{tabular} \\ \cline{2-5} 
                                                                                       & \begin{tabular}[c]{@{}l@{}}Arithmetic\\ Check\end{tabular}               & -                                                                            & \begin{tabular}[c]{@{}l@{}}Check runtime values against\\ arithmetic constraints.\end{tabular}                                                                  & \begin{tabular}[c]{@{}l@{}} \lstinline|require(b != 0); c = a / b;|\end{tabular} \\ \cline{2-5} 
                                                                                       & \begin{tabular}[c]{@{}l@{}}Address \\ Validity\\ Check\end{tabular}                  & -                                                                            & \begin{tabular}[c]{@{}l@{}}Check whether a contract address \\ is a valid one.\end{tabular}                                                                                               & \begin{tabular}[c]{@{}l@{}} \begin{tabular}[c]{@{}l@{}} \lstinline|require(owner != address(0));|                                                                             \end{tabular} \end{tabular} \\ \cline{1-5} 
    \end{tabular}%
    }
\end{table*}

\vspace{0.5em}
\textbf{Finding 3: }\textit{Transaction-reverting statements are commonly used to perform seven types of authority verifications or validity checks.}
\vspace{0.5em}

\textbf{Authority Verification.}
76 of the 435 clauses in the 382 transaction-reverting statements in the dapp contracts are for \textit{Authority Verification}. The figure for the template contracts is 31 of 175. Authority Verification aims to check whether a given contract address or token ID is authorized by the contract owner for the sake of security:

\begin{itemize}[nosep, wide]
    \item \textit{\textbf{Address Authority Check}} is to check whether a given address, mostly the address of the transaction sender, is authorized by the contract owner. We observed two types of address checks. One is to check whether the given address equals to a specified address. The other is to check whether the given address is within a list of authorized addresses. The proportions of transaction-reverting statements that perform address authority checks in dapp contracts and template contracts are 14.9\% and 16.0\%, respectively.
    \item \textit{\textbf{Token Verification}}. Tokens are value counters stored in a contract, which are mappings of addresses to account balances.
     Token verification checks whether a given token ID is authorized, i.e.,  within the mappings of addresses. The proportions of transaction-reverting statements that perform token verification in dapp contracts and template contracts are 2.5\% and 1.7\%, respectively.
\end{itemize}

\textbf{Validity Check.} 359 of the 435 clauses in the 382 transaction-reverting statements in the dapp contracts are for validity checks. The figure for the template contracts is 144 of 175. Generally, validity checks are performed to check if certain runtime values are valid, i.e., satisfying pre-defined conditions. We observed five sub-categories of validity checks:

\begin{itemize}[nosep, wide]
    \item \textit{\textbf{Logic Check}} refers to the use of logical operators to check the validity of certain runtime values. Such checks are commonly seen in the conditions of transaction-reverting statements, such as checking the return value of a low-level function call, checking the value of a boolean flag, and so on. 37.7\% dapp contracts and 29.1\% template contracts contain transaction-reverting statements for logic checks.
    \item \textit{\textbf{Range Check}} is to check whether a runtime value (e.g., an input) is within a specific range. 29.4\% dapp contracts and 26.9\% template contracts contain transaction-reverting statements for range checks.
    \item \textbf{\textbf{Overflow/Underflow Check}} is to check whether an input value crosses the limit of the prescribed size for a data type. 14.9\% template contracts contain transaction-reverting statements for overflow/underflow checks, while the ratio is only 3.9\% for dapp contracts. We further investigated the corresponding template contracts and found that many of them adopt the \texttt{SafeMath} library, which provides safe number operations to protect contracts from overflow/underflow vulnerabilities. This also shows that template contracts emphasize more on security, comparing than ordinary dapp contracts.
    \item \textit{\textbf{Arithmetic Check}} is to check whether the value of a variable violates common constraints in arithmetic operations, such as divided by 0, mod 0, etc. These checks are not frequently performed comparing to the above categories. Only 0.7\% dapp contracts and 2.3\% template contracts contain transaction-reverting statements for \textit{arithmetic checks}. 
    \item \textit{\textbf{Address Validity Check}} is to check whether a contract address is valid. Note that this is different from the address authority check discussed above, which is to check whether an address is an authorized one (a valid address may not be authorized). For example, a common address validity check is to check whether a contract address is equal to \texttt{address(0)} in an ether transfer function. When the address is zero, a new contract will be created instead of transferring ether. To avoid such cases, address validity checks should be performed. 10.8\% dapp contracts and 7.4\% template contracts contain transaction-reverting statements for address validity checks.
\end{itemize}
	
During our manual analysis, three clauses could not be categorized into the above sub-categories. Since they are not common, we do not further discuss them in the paper.
	
From the above analysis, we can see that dapp contracts and template contracts show differences in using transaction-reverting statements.
14.9$\%$ template contracts contain transaction-reverting statements for overflow/underflow checks, while the percentage in dapp contracts is only 3.9$\%$. 
Besides, dapp contracts show higher percentages in using transaction-reverting statements for logic checks, range checks, and address validity checks. 
One possible reason is that developers will consider more specific factors when applying smart contracts to a real Ethereum environment, which could be complicated since a smart contract may need to interact with other contracts and user accounts.
Comparatively, developers should consider general factors when developing template contracts and cannot anticipate the specific conditions that may arise in real environments.

\vspace{0.5em}
\noindent 
\setlength{\fboxsep}{0.5em} 
\fbox{\parbox{0.95\linewidth}{ 
\textbf{Answer to RQ2:} 
\textit{Transaction-reverting statements are commonly used to perform authority verifications and validity checks, many of which involve security-critical constraints. Template contracts and dapp contracts have different purposes for using transaction-reverting statements.}

\vspace{0.5em}
\textbf{Implication:}
\textit{Since transaction-reverting statements often check the runtime status of smart contracts against security-critical constraints, it is crucial to ensure the proper use of such statements. Future research can study the vulnerabilities induced by various misuses of transaction-reverting statements and propose detection or repairing techniques to combat such vulnerabilities.
}
}}


\begin{table*}[]

\caption{Statistics of The Purposes of Transaction-Reverting Statements in Dapp Contracts and Template Contracts}
\label{purpose_stat}

\small

\centering
\begin{tabular}{|l|l|r|r|r|r|}
\hline
\textbf{Category}                                & \textbf{First-Level Category}        & \textbf{\# Dapp Contracts} & \textbf{Ratio}   & \textbf{\# Template Contracts} & \textbf{Ratio}   \\ \hline
\multirow{2}{*}{Authority Verification} & Address Authority Check               & 65                  & 14.9\%        & 28                    & 16.0\%  \\ \cline{2-6} 
                                        & Token Verification          & 11                  &  2.5\%       & 3                     & 1.7\%   \\ \hline
\multirow{6}{*}{Validity Check}         & Logic Check                 & 164                  & \textbf{37.7}\%        & 51                    & \textbf{29.1}\%  \\ \cline{2-6} 
                                        & Range Check                 & 128                  & 29.4\%        & 47                    & 26.9\%  \\ \cline{2-6} 
                                        & Overflow/Underflow Check    & 17                  & \textbf{3.9}\%        & 26                    & \textbf{14.9}\%  \\ \cline{2-6} 
                                        & Arithmetic Check & 3                  &  0.7\%       & 4                     & 2.3\%   \\ \cline{2-6} 
                                        & Address Validity Check               & 47                  & 10.8\%        & 13                    & 7.4\%   \\ \cline{2-6} 
                                        & Other                       & 0                  &  0.0\%       & 3                     & 1.7\%   \\ \hline
\multicolumn{2}{|l|}{Total}                                           & 435               & 100.0\% & 175                   & 100.0\% \\ \hline
\end{tabular}
\end{table*}

\subsection{RQ3 (Developer Customization)} \label{subsec:rq3}

\textbf{Study Methodology:}
As discussed earlier, many developers customize template contracts to develop their own smart contracts.
In RQ3, we aim to
understand how developers customize transaction-reverting statements in template contracts.

\textbf{Step 1: Mapping Template \& Custom Contracts:}
To answer RQ3, the first step is to build a dataset containing template contracts and their corresponding custom contracts.
However, real-world smart contracts rarely explicitly specify whether they are customized from a certain template or not.
To address this problem, we leveraged code clone detection techniques to compute the similarities between each of our collected template contracts and the dapp contracts.
We considered a dapp contract customized from a template contract if the two contracts have a high similarity.

A smart contract can contain multiple components, including contracts, libraries, and interfaces. 
In practice, different components are usually put in one file in dapp contracts, while in template contracts, a file usually contains a single component.
To normalize the two kinds of contracts, we first broke down the dapp contracts into components and then compared the contracts at the component level.
Table~\ref{data pre-processing} presents the result after this pre-processing step, where \#~Contracts, \#~Interfaces, and \#~Libraries represent the number of individual contracts, libraries, and interfaces, respectively.
\begin{table}[]

\caption{Contract Pre-Processing Result}
\label{data pre-processing}

\centering
\begin{tabular}{lrrrr}
\hline
                     & \begin{tabular}[c]{@{}l@{}} \textbf{\# Contracts}\end{tabular}  & \begin{tabular}[c]{@{}l@{}}\textbf{\# Interfaces}\end{tabular}  & \begin{tabular}[c]{@{}l@{}}\textbf{\# Libraries} \end{tabular}\\ \hline
Template contracts (270)         & 190                    & 32                      & 27                    \\ 
Dapp contracts (3,866)         & 4,563       & 1,191                  & 484                   \\ 
\hline
\end{tabular}%
\end{table}


    
    

We then adopted a code clone detector, SmartEmbed~\cite{gao2020checking}, to compare the dapp contracts with the template contracts.
SmartEmbed computes similarities between two contracts based on word embeddings, and it has been shown to be effective in code clone detection for smart contracts.
Following the original experiment setting of SmartEmbed, a dapp contract is considered a custom contract of a template contract if the similarity between these two contracts is higher than 85\%. 
We chose this threshold because it achieves the highest recall in the evaluation of SmartEmbed.

As interfaces do not contain any statements, we excluded them from our dataset. 
In total, we obtained 175 contracts and 52 libraries from dapp contracts that are similar to template contracts.
These contracts and libraries form the custom contracts dataset for our subsequent analysis. 
\begin{table*}[]
  \centering
  \caption{Customization Patterns of Transaction-Reverting Statements}
  \label{require_fre_patterns}
  \resizebox{\textwidth}{!}{%
  \begin{tabular}{|l|l|l|r|r|}
  \hline
  \textbf{Category}       & \textbf{Customization Patterns} & \textbf{Description}                                                                                                                                                                                                            & \textbf{Count} & \textbf{Percentage} \\ \hline
  \multirow{4}{*}{Add}    & Add Clauses*                    & Add new clauses to a condition.                                                                                                                                                                                                 & 47             & 8.9\%               \\ \cline{2-5} 
                          & Add Variables*              & Use new variables in the condition.                                                                                                                                                                                             & 27             & 5.1\%               \\ \cline{2-5} 
                          & Add Statements                  & Add new transaction-reverting statements.                                                                                                                                                                                       & \textbf{161}   & \textbf{30.4\%}     \\ \cline{2-5} 
                          & \multicolumn{2}{l|}{Sub-Total}                                                                                                                                                                                                                                    & \textbf{235}   & \textbf{44.4\%}     \\ \hline
  \multirow{4}{*}{Delete} & Delete Clauses*                 & Remove some clauses from a condition.                                                                                                                                                                                           & 41             & 7.8\%               \\ \cline{2-5} 
                          & Delete Variables*               & Remove some variables used in the condition.                                                                                                                                                                                    & 24             & 4.5\%               \\ \cline{2-5} 
                          & Delete Statements               & Delete some transaction-reverting statements.                                                                                                                                                                                    & \textbf{104}   & \textbf{19.7}\%     \\ \cline{2-5} 
                          & \multicolumn{2}{l|}{Sub-Total}                                                                                                                                                                                                                                    & 169            & 31.9\%              \\ \hline
  \multirow{3}{*}{Change} & Modify Statement Types      & \begin{tabular}[c]{@{}l@{}}Change transaction-reverting statements to other kinds \\ of statements (e.g., change a \texttt{require} statement \\ to an \texttt{if} statement) while the condition remains the same.\end{tabular} & 4              & 0.8\%               \\ \cline{2-5} 
                          & Modify Clauses                  & \begin{tabular}[c]{@{}l@{}}Make modifications to the clauses in the condition \\ of a transaction-reverting statements.\end{tabular}                                                                                            & 59             & 11.2\%              \\ \cline{2-5} 
                          & \multicolumn{2}{l|}{Sub-Total}                                                                                                                                                                                                                                    & 63             & 11.9\%              \\ \hline
  Other                   & Cosmetic Changes                & \begin{tabular}[c]{@{}l@{}}Modifications that do not change the semantics \\ of the transaction-reverting statements.\end{tabular}                                                                                              & 62             & 11.7\%              \\ \hline
  \multicolumn{3}{|l|}{Total}                                                                                                                                                                                                                                                                 & 529            & 100.0\%             \\ \hline
  \end{tabular}%
  }
  \begin{tablenotes}
      \item[1]Note: Patterns marked with * are defined by an existing work~\cite{pan2009toward}. The remaining patterns are newly identified by us. 
    \end{tablenotes}
  \end{table*}
\textbf{Step 2: Detecting Customization Patterns:} We leveraged the custom contracts to investigate how developers customized transaction-reverting statements.
Inspired by an existing study that characterizes changes to \texttt{if} statements~\cite{pan2009toward}, we derived a taxonomy of possible customization patterns for transaction-reverting statements as shown in Table~\ref{require_fre_patterns}.
Since transaction-reverting statements are also conditional statements, the change patterns of \texttt{if} conditional statements can also be applied to transaction-reverting statements.
However, we found that patterns identified in the existing work (marked with * in Table~\ref{require_fre_patterns}) are not sufficient to cover all customizations on transaction-reverting statements.
To identify more patterns, we manually analyzed 30$\%$ of the customized transaction-reverting statements.
Specifically, we compared the transaction-reverting statements in the custom contracts with those in the corresponding template contracts and identified common customizations by checking the conditions of the transaction-reverting statements.
Via the sampling and manual analysis, we identified five more patterns.

To investigate the prevalence of the customization patterns and identify commonly used ones, we implemented a static analyzer based on a Solidity parser~\cite{PythonSolidityparser} to automatically identify the occurrences of each customization pattern.
For each pair of a template contract and a corresponding custom contract, we matched their functions by the function names and input parameters.
Functions with the same name and input parameters are seen as matched function pairs.
For each matched function pair, the analyzer 1) parses the source code into AST trees, 2) extracts all transaction-reverting statements, and 3) recognizes the customization patterns by comparing the syntactic differences of the transaction-reverting statements in the two functions. More details about the analyzer can be found on our project website~\cite{dataset}.

\vspace{0.5em}
\textbf{Finding 4: } \textit{Transaction-reverting statements in template contracts are commonly customized. Developers are most likely to strengthen transaction-reverting statements by adding clauses, variables, or new transaction-reverting statements.}
\vspace{0.5em}

Table~\ref{require_fre_patterns} shows the frequency of each customization pattern. 
In the 175 custom contracts and 53 custom libraries, our analyzer identified 529 occurrences of customization patterns.
This indicates that developers commonly customize transaction-reverting statements in template contracts.
50.1\% of the customizations are statement-level changes involving the addition or removal of a transaction-reverting statement.
 27.8\%  lie in clause granularity, including adding, deleting, and modifying a clause within the condition of a transaction-reverting statement. 
It is infrequent for developers to change transaction-reverting statements to other kinds of statements while keeping the condition unchanged (0.8\%). 
In our dataset, all of the four cases in this category are changing transaction-reverting statements to general-purpose \texttt{if} statements. 

44.4\% of the customizations fall into the ``add'' category, 31.9\% fall into the ``delete'' category, and 11.9\% involved changes. Besides, 11.7\% customizations are cosmetic changes, such as changing the order of clauses, adding or removing a string message, etc.
These changes do not alter the semantics of the original transaction-reverting statements.
These results show that developers are more likely to strengthen the transaction-reverting statements in template contracts.

\vspace{0.5em}
\textbf{Finding 5: }\textit{The customized transaction-reverting statements are commonly used for range checks and logic checks.}
\vspace{0.5em}

We further investigated the purposes of customizing transaction-reverting statements. We randomly sampled 100 customized transaction-reverting statements and manually analyzed their purpose according to the taxonomy in Table~\ref{purpose of use}.

\begin{table*}[]

\caption{The Number of Use purpose for Customization Patterns of Transaction-Reverting Statements}
\label{change_purpose}

\large

\resizebox{\textwidth}{!}{%
\begin{tabular}{|l|l|r|r|r|r|r|r|r|r|r|r|}
\hline
\multirow{2}{*}{\textbf{Category}} &
    \multirow{2}{*}{\textbf{Sub-Category}} &
    \multirow{2}{*}{\textbf{\# Total}} &
    \multicolumn{3}{c|}{\textbf{Add}} &
    \multicolumn{3}{c|}{\textbf{Delete}} &
    \multicolumn{2}{c|}{\textbf{Modify}} &
     \multicolumn{1}{c|}{\textbf{Other}} \\ \cline{4-12} 
    &
    &
    &
    \begin{tabular}[c]{@{}l@{}}\textbf{\# Add}\\ \textbf{Statements}\end{tabular} &
    \begin{tabular}[c]{@{}l@{}}\textbf{\# Add} \\ \textbf{Clauses}\end{tabular} &
    \begin{tabular}[c]{@{}l@{}}\textbf{\# Add} \\ \textbf{Variables}\end{tabular} &
    \begin{tabular}[c]{@{}l@{}}\textbf{\# Delete} \\ \textbf{Statements}\end{tabular} &
    \begin{tabular}[c]{@{}l@{}}\textbf{\# Delete}\\ \textbf{Clauses}\end{tabular} &
    \begin{tabular}[c]{@{}l@{}}\textbf{\# Delete}\\ \textbf{Variables}\end{tabular} &
    \begin{tabular}[c]{@{}l@{}}\textbf{\# Modify} \\ \textbf{Statement Types}\end{tabular} &
    \begin{tabular}[c]{@{}l@{}}\textbf{\# Modify}\\ \textbf{Clauses}\end{tabular} &
    \begin{tabular}[c]{@{}l@{}}\textbf{\# Cosmetic} \\ \textbf{Changes}\end{tabular} \\ \hline
\multirow{2}{*}{\begin{tabular}[c]{@{}l@{}}Authority \\ Verification\end{tabular}} &
    Address Authority Check &
    17 &
    10 &
    0 &
    0 &
    4 &
    0 &
    0 &
    0 &
    2 &
    1 \\ \cline{2-12} 
    &
    Token Verification &
    0 &
    0 &
    0 &
    0 &
    0 &
    0 &
    0 &
    0 &
    0 &
    0 \\ \hline
\multirow{6}{*}{\begin{tabular}[c]{@{}l@{}}Validity \\ Check\end{tabular}} &
    Logic Check &
    \textbf{34} &
    \textbf{19} &
    0 &
    0 &
    \textbf{8} &
    0 &
    0 &
    0 &
    1 &
    6 \\ \cline{2-12} 
    &
    Range Check &
    \textbf{54} &
    \textbf{27} &
    1 &
    4 &
    \textbf{5} &
    3 &
    2 &
    0 &
    6 &
    6 \\ \cline{2-12} 
    &
    \begin{tabular}[c]{@{}l@{}}Overflow/\\ Underflow Check\end{tabular} &
    0 &
    0 &
    0 &
    0 &
    0 &
    0 &
    0 &
    0 &
    0 &
    0 \\ \cline{2-12} 
    &
    Arithmetic Check &
    0 &
    0 &
    0 &
    0 &
    0 &
    0 &
    0 &
    0 &
    0 &
    0 \\ \cline{2-12} 
    &
    Address Validity Check &
    7 &
    3 &
    0 &
    1 &
    1 &
    0 &
    0 &
    0 &
    0 &
    2 \\ \cline{2-12} 
    &
    Other &
    0 &
    0 &
    0 &
    0 &
    0 &
    0 &
    0 &
    0 &
    0 &
    0 \\ \hline
\multicolumn{2}{|l|}{Total} &
    112 &
    59 &
    1 &
    5 &
    18 &
    3 &
    2 &
    0 &
    9 &
    15 \\ \hline
\end{tabular}
}
\end{table*}

\begin{figure}[tb]
    \begin{center}
    \includegraphics[scale=0.96, trim=2.5cm 16.3cm 4.3cm 0cm]{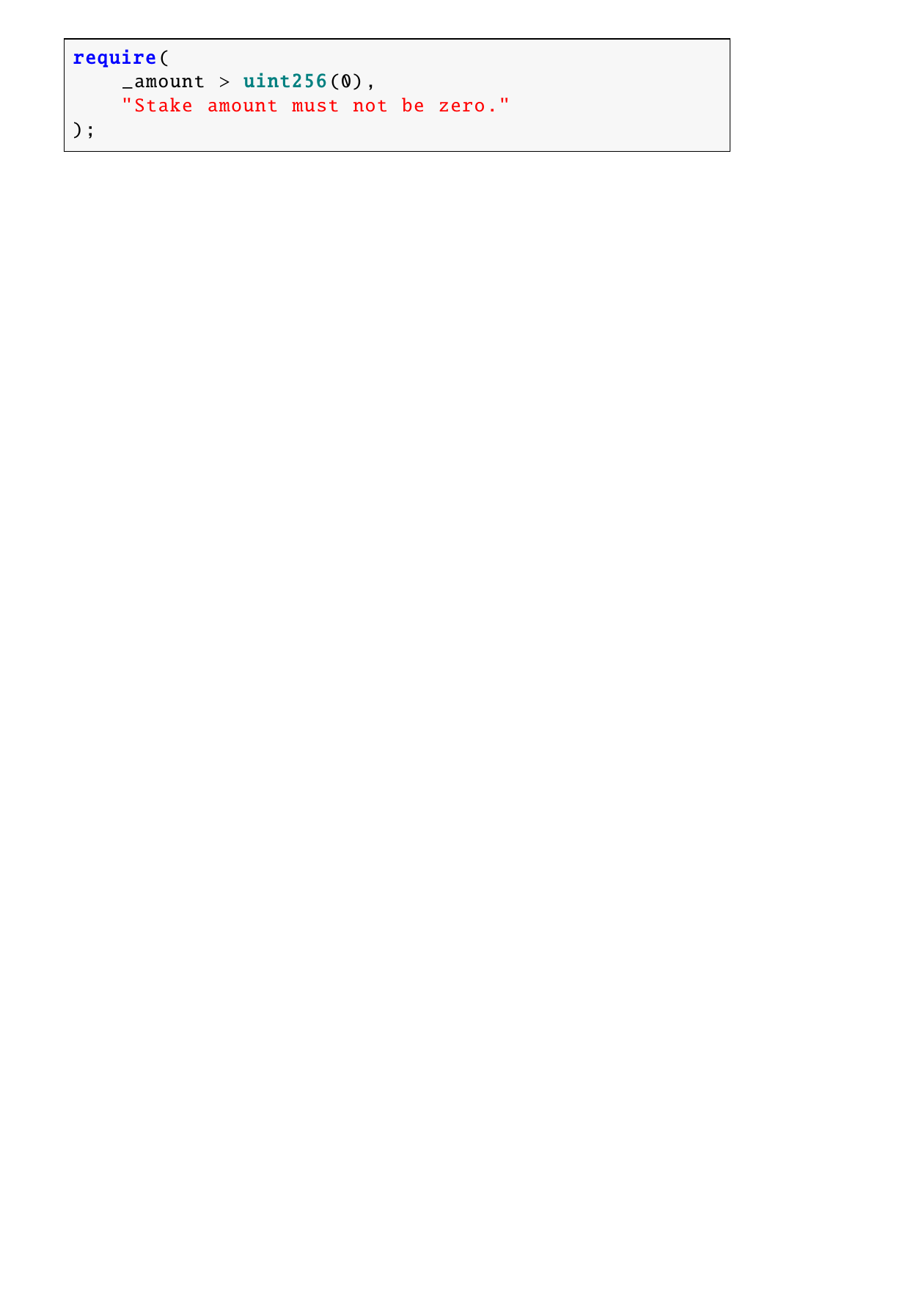}
    \caption{An example customization with \textit{Range Check} purpose}
    \label{range-check-example}
    \end{center}
\end{figure}

Table~\ref{change_purpose} shows the analysis results.
For the 100 statements, we identified 112 customized clauses and categorized them accordingly.
The results indicate that the most frequent purposes of the customized transaction-reverting statements are \textit{logic check} (30.4\%), \textit{range check} (48.2\%), and \textit{address validity check} (6.3\%).
This is reasonable since custom contracts may need to deal with use cases different from those encountered by template contracts.
They would naturally have different definitions of the validity of runtime values.
Figure~\ref{range-check-example} shows an example of a transaction-reverting statement in a custom contract.

It is a stake contract that allows EIP20 token to be staked. Staking is the process of investing tokens into the network and get a reward for doing it.
Compared with the EIP20 token template contract, the custom contract adds a transaction-reverting statement to do \textit{Range Check} to ensure that the staked amount provided by a staker is greater than 0, which intends to prevent the \textit{Integer Underflow} vulnerability~\cite{swc101}.

\begin{figure}[tb]
    \begin{center}
    \includegraphics[scale=0.96, trim=2.5cm 16.6cm 4.3cm 0cm]{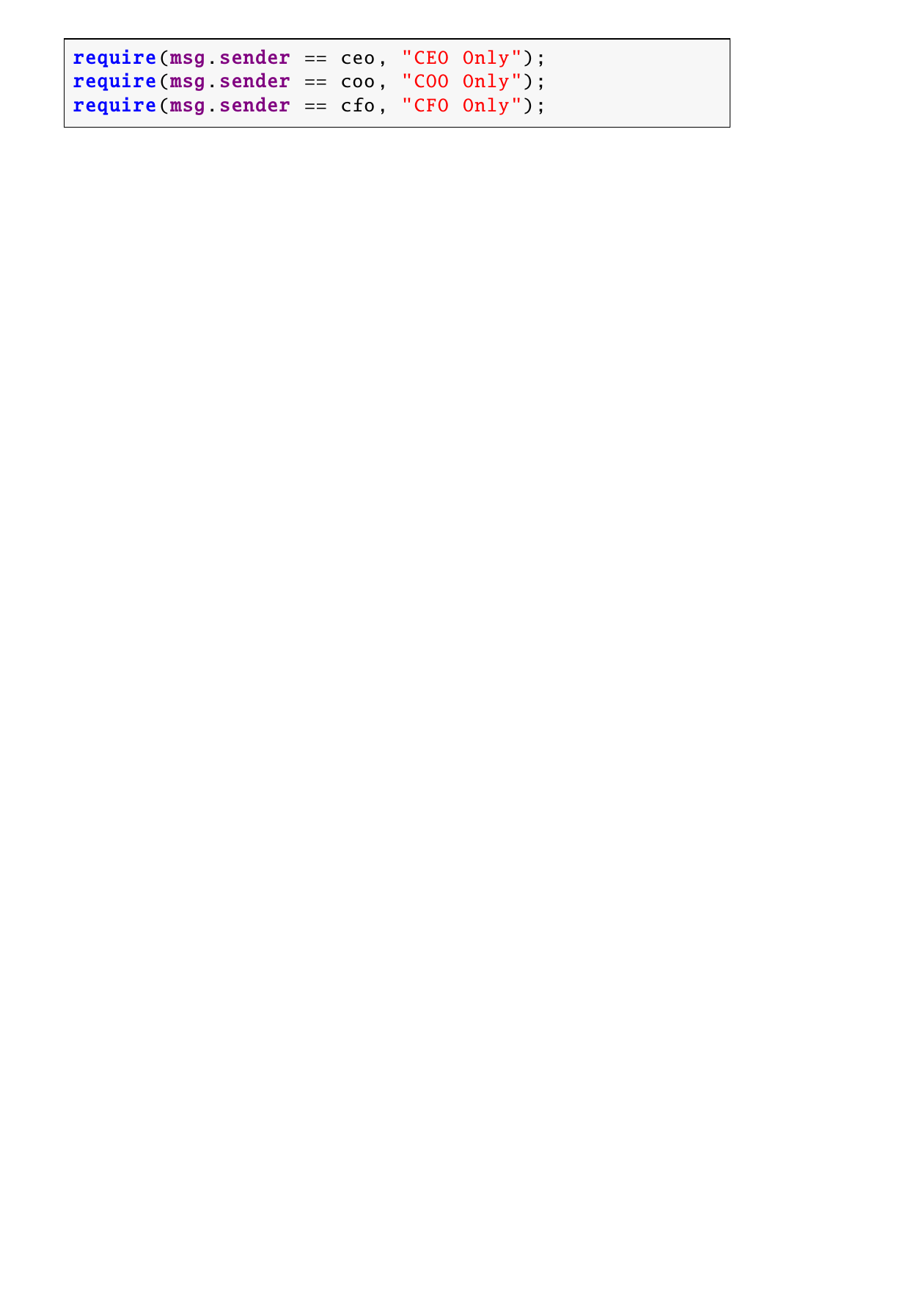}
    \caption{An example customization with \textit{Address Authority Check} purpose}
    \label{address-check-example}
    \end{center}
\end{figure}

The other 17 (15.2\%) customizations are related to \textit{address authority check}, among which, ten added statements to perform authorization check on addresses and four deleted statements for address authority check. 
This is also understandable since custom contracts can have customized permission settings for different account types.
For example, if there are multiple authorized users with different identities, the custom contract should add new transaction-reverting statements to verify the identity of the transaction sender to prevent unauthorized operations, as shown in Figure~\ref{address-check-example}.
 
\noindent 
\setlength{\fboxsep}{0.5em} 
\fbox{\parbox{0.95\linewidth}{ 
\textbf{Answer to RQ3:} 
\textit{Transaction-reverting statements in template contracts are commonly customized when developing smart contracts. Developers tend to strengthen transaction-reverting statements, mainly for logic and range checks.}

\vspace{0.5em}
\textbf{Implication:}
\textit{The customizations of transaction-reverting statements often serve security purposes. Future research may also focus on investigating the security impact of the customizations of transaction-reverting statements.}
}}

\vspace{0.5em}

\subsection{RQ4 (Security Impact)} \label{subsec:rq4}






\textbf{Study Methodology:}
To answer RQ4, we mutated the dapp contracts by removing the transaction-reverting statements.
We then leveraged smart contract security analyzers to detect the vulnerabilities in the original and the mutated contracts and compared the detection results.
Specifically, we adopted a state-of-the-art framework, SmartBugs~\cite{durieux2020empirical}, to conduct the study.
It integrates nine smart contract security analyzers, including HoneyBadger~\cite{torres2019art}, Slither~\cite{Feist_2019}, Manticore~\cite{mossberg2019manticore}, etc.
Collectively, these nine analyzers can detect 141 types of vulnerabilities, while many of them refer to the same types of vulnerabilities but have different names. 
To unify the vulnerability types, we followed the existing practices~\cite{durieux2020empirical} and used DASP~\cite{DASP}, a smart contract vulnerability taxonomy, to categorize the reported vulnerabilities. 
Another problem with these analyzers is that they may generate many false alarms due to the imprecise static analyses~\cite{durieux2020empirical}.
To mitigate this problem, we followed the existing practice~\cite{durieux2020empirical} and only counted those vulnerabilities that are reported by at least two of the nine analyzers.

\vspace{0.5em}
\textbf{Finding 6: }\textit{Missing transaction-reverting statements can introduce security vulnerabilities to smart contracts. }
\vspace{0.5em}



\begin{table}[]

\centering
\begin{threeparttable}[ht]
    \caption{The Numeber of Contracts With at Least One Vulnerability Detected by Multiple Analyzers} 
    \label{voting_result}

    \begin{tabular}{lrrr}
    \hline
    \textbf{Vulnerability Category} & \textbf{\# Before} & \textbf{\# After} & \textbf{Increase Ratio} \\ \hline
    Access Control                                                    & 3,857      & 3,857     & 0.00\%                                                    \\ 
    Arithmetic                                                        & 1,871      & 2,041     & 9.09\%                                                    \\ 
    Denial Service                                                    & 174       & 178      & 2.30\%                                                    \\ 
    Reentrancy                                                        & 634       & 671      & 5.84\%                                                    \\ 
    Unchecked Low Calls                                               & 254       & 255      & 0.39\%                                                    \\ 
    Front Running                                                     & 550       & 621      & \textbf{12.90}\%                                                   \\ 
    Time Manipulation                                                 & 132       & 159      & \textbf{16.98}\%                                                   \\ 
    Unknown Unknowns                                                  & 690       & 738      & 6.96\%                                                    \\ \hline
    \end{tabular}   

    ``\# Before'' and ``\# After'' present the security vulnerability detection results for the original contracts and the mutated contracts, respectively.
\end{threeparttable}
\end{table}


Table~\ref{voting_result} presents the number of original dapp contracts and mutated contracts that are reported to contain vulnerabilities. 
The number of vulnerable contracts increases after removing the transaction-reverting statements.
In particular, the number of contracts containing \textit{Time Manipulation} and \textit{Front Running} vulnerabilities increase significantly, by 16.98\% and 12.90\%, respectively.
This shows that transaction-reverting statements are useful in improving the security of smart contracts.
To ease understanding, we provide an example.
\begin{figure}[tb]
    \begin{center}
    \includegraphics[width=\linewidth]{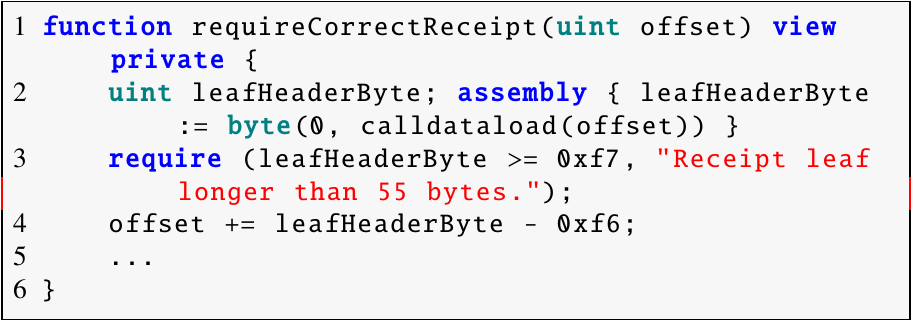}
    \caption{The number of vulnerabilities increases after mutation in contract 0x9F91b5Aa41b9fbDae6877593910586484d291F05.}
    \label{fig:vul-increase}
    \end{center}
\end{figure}
Figure~\ref{fig:vul-increase} shows a code snippet in a real smart contract. 
After removing the transaction-reverting statement in Line~3, the contract is reported to have an \textit{Underflow/Overflow} vulnerability~\cite{swc101}. 
In this example, both \texttt{leafHeaderByte} and \texttt{offset} are unsigned integers.
If Line~3 is removed, the value of \texttt{leafHeaderByte} in Line~4 can be smaller than \texttt{0xf7}, which may lead to an underflow.



\vspace{0.5em}
\textbf{Finding 7: }\textit{Smart contract security analyzers can fail to analyze transaction-reverting statements properly and induce false negatives in security vulnerability detection. 
}
\vspace{0.5em}

When inspecting the results reported by the nine analyzers, we found that there are also cases where vulnerabilities in original smart contracts disappear after removing transaction-reverting statements.
This is counter-intuitive as we have found that transaction-reverting statements are commonly used for security checks.
We found out that eight out of the nine contract analyzers (except Maian~\cite{maian}) used in our study suffered from this problem.
We further inspected such cases identified in our dataset. 

\begin{figure}[tb]
	\begin{center}
    \includegraphics[width=\linewidth]{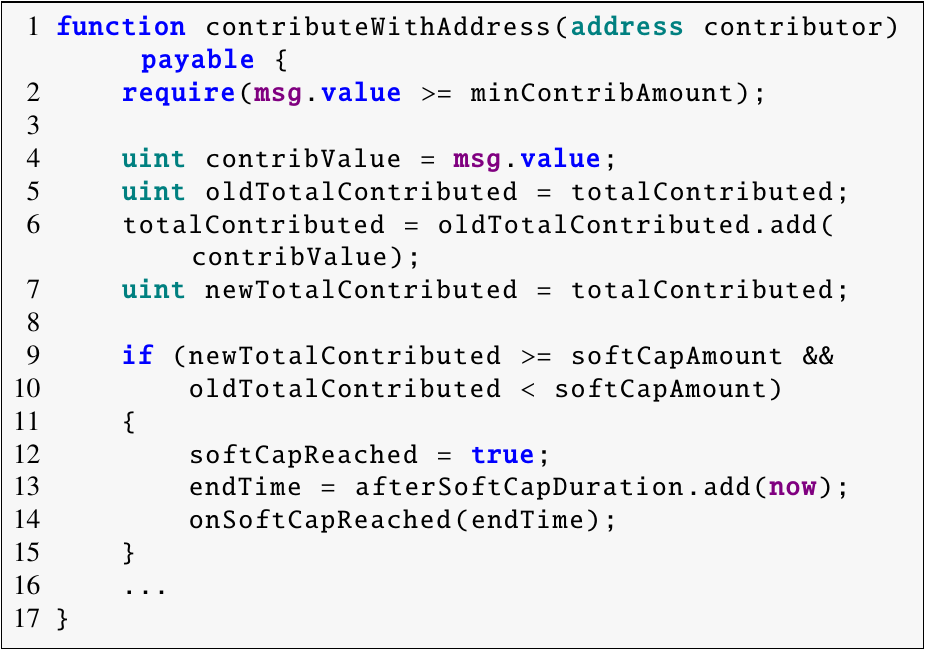}
    \caption{The number of vulnerabilities decreases after mutation in contract  0x0AbdAce70D3790235af448C88547603b945604ea.}
    \label{fig:vul-decrease}
    \end{center}
\end{figure}

Figure~\ref{fig:vul-decrease} shows a code snippet in a real smart contract.
The function \texttt{contributeWithAddress()} is reported to have a \textit{Timestamp Dependence} vulnerability~\cite{swc116}.
Due to the direct use of \texttt{now} (Line~16) which is an alias of \texttt{block.timestamp}, a malicious block miner can manipulate the block's timestamp to gain profits from the contract. 
In this case, the transaction-reverting statement in Line~2 is not related to the vulnerability as it is not checking against block timestamp.
In other words, after removing it, the \textit{Timestamp Dependence} vulnerability should still exist.
However, the tool Osiris~\cite{torres2018osiris} does not report the vulnerability after removing this transaction-reverting statement. 
This shows that Osiris can be fooled by the removal of transaction-reverting statements and induce false negatives.
We observed 5,404 such cases in our dataset where the originally detected vulnerabilities disappeared after removing transaction-reverting statements.
In our future work, we plan to take a deeper look into this problem and investigate why removing transaction-reverting statements can fool smart contract security analyzers.

\vspace{0.5em}
\noindent 
\setlength{\fboxsep}{0.5em} 
\fbox{\parbox{0.95\linewidth}{ 
\textbf{Answer to RQ4:} 
\textit{Missing transaction-reverting statements can induce security vulnerabilities in smart contracts. In other words, transaction-reverting statements can be used to avoid vulnerabilities effectively. However, there are also cases where removing irrelevant transaction-reverting statements can fool smart contract analyzers and induce false negatives in security vulnerability detection.}

\vspace{0.5em}
\textbf{Implication:}
\textit{Researchers need to further improve the effectiveness of smart contract security analyzers. In particular, properly dealing with transaction-reverting statements is a basic and critical requirement for such tools.
}
}}
\vspace{0.5em}

\section{Threats to validity}\label{sec:threats}
The validity of our study results may be subject to several threats. 
First, our selected template contracts may not be sufficiently diverse or representative. To mitigate this threat, we considered the popularity of the templates in the selection process. The four template contract repositories are all widely used by developers on GitHub~\cite{github}.
Second, we proposed a taxonomy to categorize the purposes of using transaction-reverting statements in Table~\ref{purpose of use}. There could be other ways to categorize the purposes.
To address this threat, we followed the widely-used open coding procedure to derive the results.
Third, our study results may be affected by human subjectivity, which is a common problem in qualitative coding~\cite{chandra2019qualitative}. To reduce this threat, we followed the common research practices on manual labeling by involving multiple people. Three authors iterated the labeling process three times to obtain the final taxonomy. This helped improve the reliability and generality of our taxonomy. Our data is also released for public scrutiny~\cite{dataset}.
Fourth, we used a code clone technique, SmartEmbed~\cite{gao2019smartembed}, to identify custom contracts of template contracts and set the code similarity threshold as 85\% following the experiments in the original paper to reduce false negatives. The chosen code clone technique and threshold may affect the mapping results. Also, the subjects used when investigating RQ3 are limited. We will keep expanding our dataset in the future and trying other clone detectors to see if more reliable results can be obtained.
Lastly, we used a framework supporting nine smart contract security analyzers to detect vulnerabilities in RQ4. False positives and false negatives can both exist in the results. To reduce the threat, we only kept results for items detected as vulnerable by more than one analyzer. 
Besides, the quality of transaction-reverting statements used in our constructed contract dataset may affect the accuracy of the results since our analysis is based on the assumption that the transaction-reverting statements analyzed are correct.
We plan to conduct more experiments and analyses in future studies to validate our findings further.



\section{Related Work}\label{sec:related_work}
\textbf{Error-handling Statements.}
Various studies have been conducted to characterize error-handling statements in other areas.
Filho et al.~\cite{castor2007extracting} studied the impacts of factors that affect the exception handling code in aspect-oriented programming (AOP) techniques.
Tian et al.~\cite{tian2017automatically} conducted a comprehensive study of error-handling bugs and their fixes and implemented \textit{ErrDoc}, a tool to diagnose and repair error-handling bugs in C programs automatically.
Some other studies~\cite{weimer2004finding, susskraut2006automatically, lawall2010finding, jana2016automatically, jia2019detecting} automatically detected and patched error-handling bugs using a variety of techniques.
Different from the previous studies, our work conducts the first empirical study on transaction-reverting statements (a type of error-handling statements) for Ethereum smart contracts.
It reveals the security impact of transaction-reverting statements, which is specific to smart contracts.

In terms of smart contracts, several previous studies have discussed the usefulness of transaction-reverting statements for providing defenses for vulnerabilities. 
Xue et al.~\cite{xue2020cross} showed that the \texttt{require} statement could be used to prevent reentrancy vulnerability. 
Zhou et al.~\cite{zhou2020ever} observed that most smart contracts implemented defenses via transaction-reverting statements to abort a transaction when noticing an attack. 
However, these studies only reported the use cases of transaction-reverting statements for specific purposes and did not regard them as their major focuses.
In comparison, our work is the first empirical study that systematically characterizes the use of transaction-reverting statements in real-world smart contracts.

\textbf{Smart Contract Vulnerability Detection.} In recent years, there have been many studies targeting smart contract vulnerability detection.
Static analysis methods inspected the code of smart contracts without executing them. Examples are \textit{Oyente}~\cite{luu2016making}, \textit{Zeus}~\cite{kalra2018zeus}, \textit{Vandal}~\cite{brent2018vandal}, \textit{Securify}~\cite{tsankov2018securify}, \textit{F* Framework}~\cite{bhargavan2016formal}, and \textit{Fether}~\cite{yang2019fether}.
Dynamic analysis methods check the runtime behavior of smart contracts to detect vulnerabilities.
Nikolic et al.~\cite{nikolic2018finding} employed inter-procedural symbolic analysis and concrete validators for detecting real security vulnerabilities.
Ting et al.~\cite{chen2020understanding} constructed three kinds of graphs to characterize major activities on Ethereum and proposed graph-based techniques to detect security issues.
While these studies proposed different techniques to detect vulnerabilities in smart contracts, none discussed the security impact of transaction-reverting statements.
Our work showed the prevalence of transaction-reverting statements and concluded the security impact of such statements.
Our findings can help improve security vulnerability detection techniques for smart contracts.

\section{Conclusion and Future Work}\label{sec:conclusion}

In this work, we present the first empirical study on transaction-reverting statements in Ethereum smart contracts. 
Through intensive analyses of 3,866 real-world smart contracts and 270 popular template contracts, we showed that transaction-reverting statements are prevalent in smart contracts.
They are often used to check the runtime status of smart contracts against security-critical constraints.
Our study characterizes the usage of transaction-reverting statements in practice and may shed light on future research in areas such as smart contract security and quality assurance.

In the future, we plan to extend our study by investigating the challenges in properly using transaction-reverting statements and identifying security issues induced by the misuse of transaction-reverting statements.
We also plan to leverage our findings to improve the security vulnerability detection techniques for smart contracts.


\section*{Acknowledgment}
This work was supported by the National Natural Science Foundation of China (Grant No. 61932021 and No. 62002125), Hong Kong RGC/GRF (Grant No. 16207120), Hong Kong RGC/RIF (Grant No. R5034-18) and Guangdong Provincial Key Laboratory (Grant No. 2020B121201001). Lili Wei was supported by the Postdoctoral Fellowship Scheme of the Hong Kong Research Grant Council.



\bibliographystyle{IEEEtran}
\balance
\bibliography{ref}

\end{document}